# A SUMMARY OF THE BEAT-WAVE EXPERIMENTS AT ECOLE POLYTECHNIQUE

F. Amiranoff[1], D. Bernard[3], B. Cros[4], F. Jacquet[3], G. Matthieussent[4], J.R. Marquès[1], P. Miné[3],
P. Mora[5], A. Modena[7], J Morillo[2], F. Moulin[1], Z. Najmudin[7], A.E. Specka[3], C. Stenz[6]
[1]LULI, Ecole Polytechnique - CNRS, 91128 Palaiseau, France
[2]SESI, CEA, Ecole Polytechnique, 91128 Palaiseau, France
[3]LPNHE, Ecole Polytechnique - IN2P3 - CNRS, 91128 Palaiseau, France
[4]LPGP, Université Paris Sud - CNRS, 91405 Orsay, France
[5]CPHT, Ecole Polytechnique - CNRS, 91128 Palaiseau, France
[6]GREMI, Université d'Orléans - CNRS 45000 Orléans, France
[7] Imperial College, Blackett Laboratory, Prince Consort Road, London, SW7 2AZ, UK

*Abstract*

In a large set of experiments we have studied the physics of particle acceleration by laser beat-wave driven plasma-waves using a Nd-glass laser with wavelengths close to 1 μm. The plasma is generated by multiphoton-ionization in deuterium. The plasma wave generated by the beat-wave has been observed by Thomson-scattering and its saturation attributed to the coupling with ion waves through the modulational instability. Electrons injected at 3 MeV have been accelerated in the plasma up to 4.3 MeV. The maximum energy gain is limited by the dephasing between the injected electrons and the plasma wave. The observed energy gain is compatible with a maximum accelerating electric field of 0.7 GV/m.

## 1 INTRODUCTION

The beating between two copropagating electromagnetic waves in a plasma can generate a longitudinal electron plasma wave with a high electric field and a relativistic phase velocity. This mechanism called beat-wave [1] is efficient if the electron plasma frequency $\omega_p$ is close to the difference frequency between the two laser beams. It is one of the proposed mechanisms for laser particle acceleration. Various experimental programs have been developed in order to study this mechanism and the possible resulting electron acceleration [2-7]. Among them, several groups have used a $CO_2$ laser at wavelengths near 10 μm [3,4,6]. We describe here the results obtained at Ecole Polytechnique with a Nd-glass laser at wavelengths near 1 μm. The different issues that were successively studied, and that we describe here, include plasma formation [8], the generation and the measurement of the relativistic electron plasma wave, the physics relevant to its saturation [9,10] and the acceleration of an injected electron beam [7].

## 2 PLASMA FORMATION

The plasma wave grows significantly when the resonance condition between the plasma frequency and the laser frequencies is fulfilled. The required precision on the plasma density depends on the laser wavelength and intensity. In absence of saturation, it is of the order of 1% to a few % respectively [11] for existing Nd-glass lasers ($\lambda \approx 1$ μm) and $CO_2$ lasers ($\lambda \approx 10$ μm). To obtain such a precision and homogeneity we generate the plasma by multiphoton-ionization of deuterium. Multiphoton ionization presents a very sharp threshold with incident laser flux. Above this threshold, the gas is fully ionized and the electron density is determined by the initial atomic density. We demonstrated this in a previous experiment by measuring the electron density by Thomson scattering [8], with a frequency doubled Nd laser beam at $\lambda \approx 0.5$μm, pulse duration 200ps (600ps) and energies up to 10J (30J). After ionization, the ponderomotive force of the laser beam pushes the electrons outwards and the ion and electron densities slowly decrease by a few percent per 100 ps.

## 3 GROWTH AND SATURATION OF THE PLASMA WAVE

All the beat wave experiments have been made with two laser beams at $\lambda \approx 1.0530$μm and $\lambda \approx 1.0642$μm, pulse duration of about 100ps and energies up to 10J per wavelength. In a first series of experiments we studied the growth and the saturation of the plasma wave generated by beat-wave. The two pulses are synchronized and focused in the middle of the chamber filled with deuterium[12]. After ionization the plasma wave begins to grow. The direct observation of this wave by collinear Thomson scattering at 0° is not reliable because of a large signal due to a four-wave coupling mechanism in the gas surrounding the plasma [10,9].

When its amplitude is high enough, the plasma wave couples with electron and ion waves, with longer $k$ vectors, in the regime of modulational instability [9,10]. These waves are measured by Thomson scattering of a collinear probe beam at $\lambda = 0.53$ μm. In this configuration, the scattering angle on the direct waves is very small. Thus, the scattered spectra observed at 10° reveal the existence of long *k* vector electron **and** ion waves. This is the evidence of an efficient coupling of the primary plasma wave with other modes in the plasma. This mechanism saturates the growth of the primary wave very early in the laser pulse at a rather low level of density perturbation of 1% to 5%. The imaging of the secondary waves for a focusing length of 1 m shows the presence of beatwave over a length of 1 cm and a maximum diameter of the order of 300 µm.

## 4 ELECTRON ACCELERATION

In the last series of experiments we studied the acceleration of a relativistic electron beam injected in the plasma. The set-up is shown in Fig.1. The two laser pulses are focused by a 1.5 m or 1.2 m focal length lens in deuterium gas. At resonance, the deuterium density is $1.115 \times 10^{17}$ cm$^{-3}$ ±2.3% corresponding to a pressure equal to 2.272 mbar at 22°C. It can be adjusted with a precision of ±0.3%.

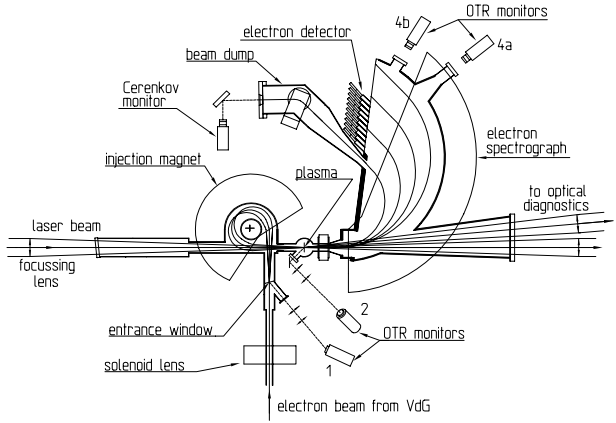

Figure 1 : Experimental set-up of the acceleration experiment.

The electron source is a pulsed Van de Graaff accelerator delivering electrons with a total energy of 2.5, 3.0 or 3.3 MeV and a relative energy fluctuation of 10$^{-3}$. The pulse duration is equal to 0.4 ms, so the beam can be considered as continuous during the life time of the plasma wave. The current is set to 170 µA corresponding to 1000 electrons per picosecond. The beam is focused on a thin (1.5 µm) aluminium foil separating the vacuum of the beam pipe from the target chamber, and imaged at the plasma location by a triple-focusing magnet [13,14]. The geometrical parameters of the beam have been measured using monitors that detect optical transition radiation [15]. We found 25 µm [RMS] for the focal spot at the plasma location in vacuum, 45 µm for the focal spot in 2 mbar D$_2$, and 10 mrad [RMS] for the angular divergence.

The energy spectrum of the electrons after passing the plasma is measured by a magnetic spectrograph and an array of 10 scintillators read by photomultipliers [13]. The noise is mainly due to electrons scattered on the gas molecules. It amounts to 5 e$^-$/ns. The photomultiplier signals being gated by 5 ns electronic gates, this noise amounts to 25 e$^-$ per channel.

In the following we will summarize the main results. A large number of shots have shown the presence of accelerated electrons. In a series of null tests, we have verified that accelerated electrons are detected only when the two laser pulses are synchronized in time and near the theoretical resonant pressure.

Optimal acceleration is expected when the phase velocity of the plasma wave is not too different from the velocity of the accelerated electrons. In our case the relativistic factor corresponding to the phase velocity is $\gamma_p = 94.5$ and the maximum injection energy corresponds to $\gamma_e = 6.5$. This means that the electrons get out of phase while travelling in the plasma. The dephasing distance is given by $l \approx \gamma_p \gamma_e^2 \lambda$, where $\lambda$ is the mean laser wavelength, and is equal to ≈ 4.2 mm. If the plasma is too long, the electrons get out of phase during acceleration and undergo acceleration and deceleration successively, thus decreasing the maximum energy gain. This latter is given to first order by [16] $\Delta\gamma = \Delta\gamma_m \times e^{-L/l_d}$ where $\Delta\gamma_m = eE_{max} \times \pi Z_R/mc^2$, $Z_R$ is the Rayleigh length of the focused laser beam, $L = 2 \times Z_R$ and $l_d = (2/\pi) \times l$. To check this effect we varied independently the plasma length $L$ and the injection energy.

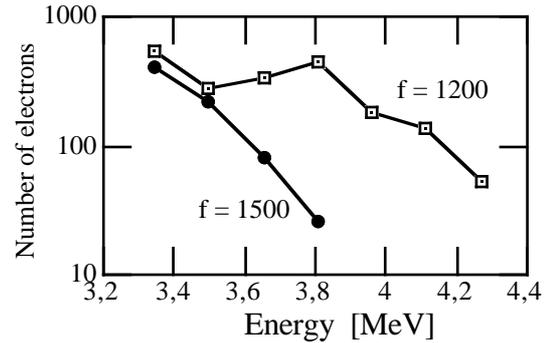

Figure 2 : Spectra of accelerated electrons obtained with a focal length of 1.5 m (black symbols) and a focal length of 1.2 m (open symbols). The injection energy was 3 MeV. The two curves show the spectra of the best shots near the resonant pressure.

Figure 2 shows the electron spectra obtained with an injection energy of 3 MeV and focal lengths of 1.5 m and 1.2 m. It appears clearly that the energy gain is much higher in the shorter plasma. Further evidence for dephasing as the limiting factor of acceleration is provided by Fig.3 where the focal length is fixed and the injection energy is varied from 2.5 MeV to 3.3 MeV. Under the same plasma conditions, the maximum energy gain increases significantly when the injection energy increases, i.e when the dephasing length increases.

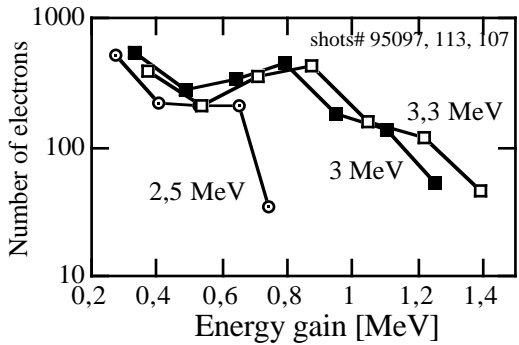

Figure 3 : Best spectra of accelerated electrons for three injection energies (2.5, 3.0 and 3.3 MeV). The focal length is 1.2 m.

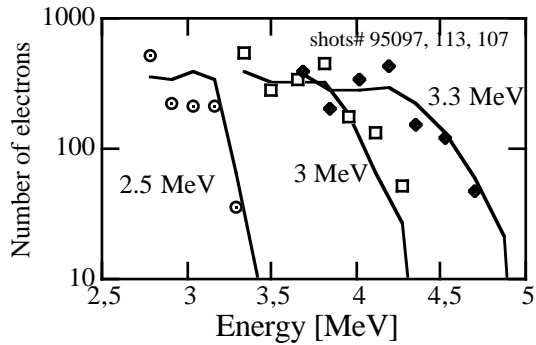

Figure 4 : Same spectra as in Fig.3. compared with the results of the acceleration model (continuous lines) for the parameters $L = 2.8$ mm and $\delta_{max} = 2.4\%$, and a plasma wave life-time of 3ps.

We compared the measured spectra with a simple 3D acceleration model [16] assuming a gaussian laser beam with the measured focal spot size and the Rayleigh length as a free parameter. The amplitude $\delta$ of the plasma wave is assumed gaussian transversally and Lorentzian longitudinally, with a peak value at focus of $\delta_{max}$. The determinations of $L$ and $\delta_{max}$ from the energy gains at two injection energies or with two different focal lengths furnish the same results, $L = 2.8$ mm and $\delta_{max} = 2.4\%$ (Fig.4), corresponding to a maximum electric field of 0.7 GV/m.

The value of the electric field is in agreement with the Thomson scattering measurements and the theoretical predictions [9]. The acceleration length of 2.8 mm is much shorter than the time-integrated image of one of the Thomson satellites (1 cm). Nevertheless, we have also seen [9] that the plasma waves only exist in a limited region at a given time because the saturation time due to the modulational instability depends on the local laser intensity. Both parameters are thus in reasonable agreement with our preceding measurements.

## 5 CONCLUSION

We have presented a summary of the results of the beatwave experiments at Ecole Polytechnique. All important physical points seem to be well understood : the growth of the plasma wave is limited by modulational instability and the energy gain of the accelerated electrons is limited by dephasing between the electrons and the plasma wave. It is quite clear from these results and the results obtained elsewhere in a different regime that future beat-wave experiments should involve more powerful laser beams with possibly shorter pulses, and injection of electrons with higher energies.

## 6 ACKNOWLEDGEMENTS

We gratefully acknowledge the help of the technical staff of the LULI, LPNHE and SESI during these experiments. This work has been partially supported by Ecole Polytechnique, IN2P3-CNRS, SPI-CNRS, CEA, EEC and DRET. We also acknowledge the support of the Human Capital and Mobility Program of the European Community (contract #CHGECT930046).

## 7 REFERENCES

[1] T. Tajima and J.M. Dawson, Phys.Rev.Lett. **43**, 267 (1979).
[2] F. Martin, J.P. Matte, H. Pepin and N.A Ebrahim, *Proceedings of the workshop on New Developments in Particle Acceleration Techniques (Orsay 1987)* edited by S. Turner (CERN, Genève, 1987), p. 360.
[3] C.E. Clayton et al., Phys. Rev. Lett. **70**, 37 (1993) ; C.E. Clayton, M.J. Everett, A. Lal, D. Gordon, K.A. Marsh, and C. Joshi, Phys. Plasmas **1**, 1753 (1994).
[4] Y. Kitagawa et al., Phys. Rev. Lett. **68**, 48 (1992).
[5] A.E Dangor, A.K.L. Dymoke-Bradshaw and A.E Dyson, Physica Scripta. **T30**, 107 (1990).
[6] N.A. Ebrahim, J. Appl. Phys. **76**, 7645 (1994).
[7] F. Amiranoff et al., Phys. Rev. Lett. **74**, 5220 (1995).
[8] J.R Marquès et al., Phys.Fluids B **5**, 597 (1993).
[9] F. Moulin et al., Phys. Plasmas **1**, 1318 (1994).
[10] F. Amiranoff et al., Phys. Rev. Lett. **68**, 3710 (1992).
[11] P. Mora and F. Amiranoff, J. Appl. Phys. **66**, 3476 (1989).
[12] F. Amiranoff et al., Rev. Sci. Instr. , **61, 2**133 (1990).
[13] F. Amiranoff et al., Nucl. Instr. and Meth. A., **363**, 497 (1995)
[14] D. Bernard, A.E. Specka, Nucl. Instr. and Meth. A. **366,** 43 (1995)
[15] A.E. Specka et al., proceedings of the IEEE Particle Accelerator Conference, pp. 2450-2452, Washington, D.C., 1993.
[16] P. Mora, J. Appl. Phys. **71**, 2087 (1992).